\documentclass[useAMS,usenatbib]{mn2e}
\usepackage{graphicx}
\usepackage{amsmath}
\usepackage{pxfonts}
\usepackage{color}

\title[Jet emitting disc spectrum]{Spectrum of jet emitting disc: Application to microquasar XTE J1118+480}
\author[Zhang and Xie]{Jian-Fu Zhang$^1$, Fu-Guo Xie$^2$\\
$^1$Department of Physics, Tongren University, Tongren 554300, China; jianfuzhang.yn@gmail.com\\
$^2$Key Laboratory for Research in Galaxies and Cosmology, Shanghai Astronomical Observatory, Chinese Academy of Sciences,\\
80 Nandan Road, Shanghai 200030, China; fgxie@shao.ac.cn}

\begin{document}
\pagerange{\pageref{firstpage}--\pageref{lastpage}} \pubyear{2013}

\maketitle

\begin{abstract}
Under the framework of the magnetized accretion ejection structures, we analyze the energy balance properties, and study the spectral energy distributions (SEDs) of the Jet Emitting Disc (JED) model for black hole X-ray transients. Various radiative processes are considered, i.e. synchrotron, bremsstrahlung, and their Comptonizations, and external Comptonization of radiation from outer thin disc. With these cooling terms taken into account, we solve the thermal equilibrium equation self-consistently and find three solutions, of which the cold and the hot solutions are stable. Subsequently we investigate the theoretical SEDs for these two stable solutions. We find the hot JED model can naturally explain the spectra of the Galactic microquasars in their hard states. As an example, we apply this model to the case of XTE J1118+480.
\end{abstract}


\section{INTRODUCTION}
Black hole X-ray transients, which are systems consisting of a stellar black hole primary accreting matter from its companion star, display fruitful and distinctive spectral and timing features during their outbursts (see \citealt{mr06,rm06,Belloni10} for reviews). Based on their X-ray properties, three states were discovered. The soft state is relatively bright, the spectrum is thermal, which can be modeled as multi-temperature blackbody radiation in the soft X-ray band. The hard X-ray flux is relatively weak in this state. The hard state, on the other hand, is usually fainter. The thermal component is very weak (if not absent), whereas the hard non-thermal X-ray radiation dominates. The intermediate state is a combination of the above two states \citep{mr06,rm06,Belloni10}. (near-)Simultaneous joint radio and X-ray observations show that the sources are also bright in radio band during their hard states (\citealt{Corbel008}). The radio emission is usually interpreted as the radiative signature of a jet, which has been resolved in a few cases (e.g., \citealt{Dhawan00,Stirling01}). High resolution radio observations of several sources reveal properties similar to quasars, i.e. two-sided jet-like structures and/or superluminal motions of the jet components, and earn them the name microquasars \citep{Mirable99}. Extensive observations have been made in recent years, which result in an accumulation of a huge amount of multi-band (from radio to ${\rm TeV}$) observational data, and lead to a better understanding of these systems.

The observations on the hard state of microquasars support the basic theoretical picture \citep{emn97} that, these systems are composed of relativistic expanding ejections (jets), plus a standard thin disc \citep[][SSD hereafter]{Sha73} which is truncated at certain radius and changes into a hot accretion flow \citep{emn97,n05}. Under this general picture, the radiation at a given waveband may be dominated by the contribution from a specified component (e.g. jet or disc), or even a specified location of one component. For example, similar to the emission of the radio cores in active galactic nuclei, the flat or slightly inverted radio (up to infrared) spectrum can be naturally interpreted as self-absorbed synchrotron radiation from relativistic jets \citep[e.g.,][]{Blandford79,Hjellming88,Esin01,Yuan05,Yuan06,Peer12}. Another example comes from the $\gamma$-ray radiation. Although under certain extreme conditions hot accretion flow can also produce strong $\gamma$-ray radiation through $\pi^0$ production from $p-p$ collisions \citep{mnk97,om03,nxs13}, the $\gamma$-ray radiation in microquasars is generally interpreted as radiation from energetic particles in relativistic jets \citep[e.g.,][]{Romero03,Romero08,Bosch06,Zhang10,Zhang11,Zdziarski12a,Zdziarski12b}.

The origin of the X-ray radiation, on the other hand, is still actively debated (see \citealt{mr06,Mar10} for recent reviews). One possibility is that the X-ray may have its origin from jets (namely the jet model), either as synchrotron \citep[e.g.,][]{Mark01,Mark03} or inverse Compton of the synchrotron (synchrotron self-Comptonization) and/or the blackbody (external Comptonization) radiation \citep[e.g.,][]{Mark05,Peer09,Peer12}. Another possibility is that the X-ray radiation is the Comptonization of synchrotron radiation within the hot accretion flow itself \citep{Esin01,Yuan05}. We note that the latter, with the name jet-(hot) accretion scenario, can also naturally explain the complicated timing features of the microquasar XTE J1118+480 \citep{Yuan05}, and makes itself more attractive compared to the jet model. In the jet-accretion model, the hot accretion flow can be the advection-dominated accretion flow (ADAF hereafter, \citealt{Nar94}) at lower luminosities or the luminous hot accretion flow (LHAF hereafter, \citealt{Yuan01}) at brighter luminosities.

One obvious caveat is that both the jet model and the jet-accretion model treat different components (e.g., jet, SSD, ADAF) in an independent way, i.e., each component can be adjusted almost freely in order to fit observations, although these components are recognized to be tightly related (We note that \citealt{xy08} investigated the dynamical impact of outflow/wind onto accretion flow.). Under the magnetized accretion ejection structures (MAES), \citet{Fer93} originally investigated the jet emitting disc model (JED hereafter, see also \citealt{Fer95,Fer97,Cas00a,Cas00b,Fer06}). Compared to the above two widely-adopted models, this new model has the advantage that it treats both the accretion disc and the ejection it generates consistently. Besides, \citet{Fer06} (see also \citealt{Petr10}) speculate that the canonical spectral states of microquasars can be promisingly explained under the JED model.

One of the important next-steps is to apply the JED model to microquasars and derive quantitatively the physical properties (e.g. accretion rate, ejection rate, etc) from spectral fitting. The first attempt in this direction was done by \cite{Foe08a,Foe08b}. The main spectral features of microquasars spectral energy distributions (SEDs) are demonstrated, and a first comparison to the observations of XTE J1118+480 is shown (see e.g. Fig.\ 3 in \citealt{Foe08a}). To go a bit further, we detail in this paper the SEDs expected from JED. Several radiative cooling processes are included in this work, i.e. synchrotron and its Comptonization, bremsstrahlung and its Comptonization, and external Comptonization. This paper is organized as follows: In the next section we present the basic properties of the JED model; numerical results are given in Section 3; in Section 4, we apply JED model to one individual source --- the microquasar XTE J1118+480; Section 5 is the summary and discussions.

\section{Model description of jet emitting disc}
As stated above, the jet emitting disc model for microquasars is developed from the MAES model. With the self-similar assumption in radial direction, it evolves vertically from the midplane of the resistive magnetohydrodynamical (MHD) disc to the ideal MHD jet. This geometric configuration is similar to the jet-accretion models \citep{Yuan05,Yuan06}, but the main difference is that accretion and ejection are treated self-consistently. In other words, the ejection efficiency is not a free parameter in a JED (compared to adiabatic inflow-outflow solutions or jet-accretion model), but results from the resolution  of the full set of dynamical MHD equations without neglecting any term. In this section, we will only present several key elements and assumptions of the model that are important to the current work. Interested readers are referred to \citet{Fer93,Fer95,Fer97,Fer02,Cas00a,Cas00b} for related details.

\subsection{Key assumptions}
In order to build the JED model, several assumptions have been made (see \citealt{Fer02,Petr10} for justifications and discussions on these assumptions):

(i) The presence of a large scale, well organized, vertical magnetic field $B_z$.

(ii) Time-independent and non-relativistic. The whole system is assumed to be dynamically steady ($\partial/\partial t=0$). Following \cite{Fer02}, we limit ourselves to the non-relativistic MHD framework.

(iii) Axisymmetric. All quantities are independent to the azimuthal angle ($\partial/\partial \phi=0$). Cylindrical coordinate is adopted.

(iv) Single-fluid MHD. The gases for accretion or ejection are likely to be ionized. For simplicity, we assume the thermal coupling between ions and electrons is fully effective, thus the JED is one-temperature. As argued later in Sec. \ref{sec:appl}, the two-temperature structure may be more realistic.

(v) Self-similarity in the radial direction. The full set of the MHD equations can be numerically solved under the self-similar approach, where none of the dynamical terms are ignored.

(vi) $\alpha$-prescription on the transport coefficients. Several transporting mechanisms are considered in this model, i.e. the magnetic diffusion (with coefficient $\nu_{\rm m}$), resistivity (with $\eta_{\rm m}$), viscosity (with $\alpha_{\rm \upsilon}$) and thermal conduction (with $\kappa_{\rm T}$). For simplicity (and also because of uncertainty in microphysics), the coefficients for these mechanisms are assumed to be the same, following the $\alpha$-prescription \citep{Sha73}.

\subsection{Radial structures}
Accretion process in SSD relies on a mechanism to transport angular momentum, which is now commonly accepted to be turbulent viscosity arising from the magneto-rotational instability (MRI) \citep{Bal91}. The MRI can only work at weak magnetic fields (i.e. the so-called magneticity $\mu = B^2_{\rm z}/\mu_{\rm o}P <1$) and will be quenched when $\mu$ approaches unity. $\mu$ may reasonably be expected to increases as the accreting material moves inward \footnote{However, recent MHD numerical simulations show that the reverse (i.e. an increase of $\mu$ with $r$) is indeed expected in a stationary state (\citealt{DeFarreira11}). Therefore, an increase of $\mu$ towards the center would suggest a non stationary situation.}(ref. Eq. 4 in \citealt{Fer06}), and it is expected that within a transition radius $r_{\rm tr}$, $\mu$ will be greater than unity and SSD will be truncated to JED. The transition radius should depend on various properties of the accreting material, e.g. the accretion rate, the magnetic diffusion, the magnetic flux, etc. With all these complexities, we take it as a free parameter throughout the current work. Within $r_{\rm tr}$, the JED accretion rate is given as
\begin{equation}
\dot{M}_{\rm a}(r) = \dot{M}_{\rm a,out} \left(\frac{r}{r_{\rm tr}}\right)^{\xi},
\label{EQmot}
\end{equation}
where $\dot{M}_{\rm a,out}$ is the accretion rate at $r_{\rm tr}$ (thus also the accretion rate for the outer SSD), and $\xi$ measures the local mass ejection efficiency (\citealt{Fer93}). With Eq. (\ref{EQmot}), the total liberated accretion power in the JED regime is
\begin{equation}
P_{\rm acc}= \frac{GM_{\rm BH}\dot{M}_{\rm a,out}}{2r_{\rm in}}\left[\left(\frac{r_{\rm in}}
{r_{\rm tr}}\right)^{\xi}-\frac{r_{\rm in}}{r_{\rm tr}}\right],
\label{eq:Pacc}
\end{equation}
where $r_{\rm in}=6~r_{\rm g}$ ($r_{\rm g}=GM_{\rm BH}/c^2$ is the gravitational radius) is the innermost radius of JED.

In the JED model, the accreting gases and ejecting gases interact with one another. Under this framework, ejections are collimated and can be called "magnetically-driven" disc wind which is the same as a self-confined jet \footnote{A fraction of the magnetic energy could be dissipated at the disc surface by the Joule effect and thereby help the launching of a wind. Thus, when thermal effects can come into play (\citealt{Cas00b}), this ejection is usually called "thermally or radiatively-driven" disc wind, which produces no significant torque and carries away only a tiny portion of the angular momentum.}. The ejections (jets) will carry away most of the angular momentum allowing the disc material to accrete (Jonathan Ferreira, private communication ). In other words, the ejection provides an additional torque onto the accretion disc to transport the angular momentum. This ejection torque (originally defined as jet torque), which is highly related to the magnetic fields, can be expressed as $t_{\rm ejec} \sim B_\varphi^+ B_{\rm z}/\mu_{\rm o} h$. Note that there is also a normal turbulent viscous torque, which following \citet{Sha73}, can be expressed as $t_{\rm disc,r\varphi}=-\alpha_\upsilon P/r$ (Here $P$ is the total pressure and $\alpha_\upsilon$ is the viscous parameter.). The ratio of the above two torques at the midplane is
\begin{equation}
\Lambda = \left|\frac{t_{\rm ejec}}{t_{\rm disc,r \varphi}}\right| \sim \frac{B_\varphi^+
B_{\rm z}/\mu_{\rm o} h}{\alpha_{\rm \upsilon} P/r}
= \frac{B_{\rm \varphi}^+ B_{\rm z}}{\mu_{\rm o} P} \frac{r}{\alpha_{\rm \upsilon} h}.\label{LMAD}
\end{equation}
The steady state ejection with $\mu\sim1$ imposes $\Lambda \sim r/h \gg 1$
\citep{Fer97,Cas00a,Fer02}. Then, the thinner the  accretion disk, the more the angular momentum is actually transported by the ejected material. The importance of the torque imposed by ejecting materials distinguishes the JED from other accretion disc models (SSD, ADAF, etc).

Now we will present the basic properties of the magnetized accretion flow in JED models. The thickness is $h(r)=c_{\rm s}r/V_{\rm K}$. Here $c_{\rm s}=(P/\rho_{\rm o})^{1/2}$ is the isothermal sound speed, where $P$ is the total (gas+radiation) pressure and $\rho_{\rm o}$ is the gas density at the equatorial plane, and $V_{\rm K}=\sqrt{GM_{\rm BH}/r}$ is the Keplerian velocity. We further define the disc aspect ratio as
\begin{equation}
\varepsilon \equiv \frac{h(r)}{r}=\frac{c_{\rm s}}{V_{\rm K}}.
\end{equation}
From Eq. (\ref{LMAD}), the total torque (jet+viscous) $t_{\rm tot}=t_{\rm ejec}+t_{\rm disc,r\phi}=t_{\rm disc,r\phi}(1+\Lambda)$. The  sonic Mach number can be deduced then from the angular momentum conservation (e.g. \citealt{Belo99}): $m_{\rm s}=-\upsilon_{\rm r}/c_{\rm s}=\alpha_{\rm \upsilon}\varepsilon(1+\Lambda)$. Obviously because of the differences in $\varepsilon$ and $\Lambda$, the SSD is subsonic ($m_{\rm s}=\alpha_{\rm \upsilon}\varepsilon\ll 1$), while the JED can be supersonic \citep{Fer95,Fer97}.

To simplify our analysis, we fix $m_{\rm s} = 1$ and the magneticity $\mu = 1$ (Note that the dynamical structure presented below can be easily generalized to other values of $m_{\rm s}$ and $\mu$.) throughout this paper. We further assume that the disc is gas-pressure supported, i.e. the radiation pressure is negligible (see also \citealt{Petr10}). The basic quantities in a JED can then be written as \citep{Fer02,Fer06,Petr10}
\begin{eqnarray}
\upsilon_{\rm r} = - m_{\rm s} V_{\rm K} \varepsilon &\simeq& -3.0\times10^{10}
\varepsilon R^{-1/2} ~\rm cm~s^{-1},\label{Uvel} 
\end{eqnarray}
\begin{eqnarray}
P=\rho_{\rm o}c_{\rm s}^2=\frac{\dot{M}_{\rm a}
V_{\rm K}}{4 \pi r^2 m_{\rm s}}&\simeq& 
1.5\times 10^{16}\dot{m}m^{-1}R^{\xi-5/2} ~\rm erg ~ cm^{-3}, \label{Pgas} 
\end{eqnarray}
\begin{eqnarray}
T = \frac{m_{\rm p}}{2k_{\rm B}}(V_{\rm K}\varepsilon)^2 &\simeq& 5.4\times 10^{12} \varepsilon^2 R^{-1} ~\rm K, \label{Temp} 
\end{eqnarray}
\begin{eqnarray}
n=\frac{\rho_{\rm o}}{m_{\rm p}} = \frac{\dot{M}_{\rm a}}{4\pi m_{\rm p}V_{\rm K} r^2 m_s
\varepsilon^2}
&\simeq&
10^{19}\varepsilon^{-2} \dot{m}m^{-1}R^{\xi-3/2}~\rm cm^{-3},
\label{numDen}
\end{eqnarray}
\begin{eqnarray}
B_{\rm z} = \sqrt{\left(\frac{\mu}{m_{\rm s}}\right)\left(\frac{\mu_{\rm o}\dot{M}_{\rm a} V_{\rm K}}{4 \pi r^2}\right)}
&\simeq& 3\times 10^{8} \dot{m}^{1/2}m^{-1/2}R^{\xi/2-5/4}~\rm G, \label{MagField}
\end{eqnarray}
where $m= M_{\rm BH}/M_{\rm \sun}$, $R= r/r_{\rm g}$ and $\dot{m} = \dot{M}_{\rm a,out}/\dot{M}_{\rm Edd}$ (Here $\dot{M}_{\rm Edd} = L_{\rm Edd}/c^2$ is the Eddington accretion rate, and $L_{\rm Edd}$ is the Eddington luminosity.).

\subsection{Global energy conservation of JED model}
The energy in a JED is globally conserved, and it can be written as
\citep{Fer93,Fer02,Com08,Petr10}
\begin{equation}
P_{\rm acc}=P_{\rm JED} +P_{\rm ejec},\label{energy}
\end{equation}
where $P_{\rm JED}$ is the power dissipated within the JED, which is the sum of the radiated power $P_{\rm rad}$ and the advected power $P_{\rm adv}$. The ejection power $P_{\rm ejec}$ is the energy flux that leaves the two disc surfaces ($P_{\rm ejec}$ is originally defined as $P_{\rm jet}$, see \citealt{Fer06,Petr10}). The ejections in our model are magnetically driven, thus their power are dominated by the magnetic part, i.e. the Poynting flux. We thus write $P_{\rm ejec}$ as
\begin{equation}
P_{\rm ejec}=2\int\vec{S}_{\rm MHD}\centerdot {\rm d}\vec{A}=-2\int_{r_{\rm in}}^{r_{\rm tr}}\Omega_{\rm m}rB^+_{\rm \varphi}B_{\rm z}
\mu_{\rm o}^{-1}2\pi r{\rm d}r.\label{pjet}
\end{equation}
Here $\Omega_{\rm m}$ is the rotational rate of the magnetic field line
on the surface \citep{Petr10}, $\vec{S}_{\rm MHD}=-\Omega_{\rm m}rB_{\rm \varphi}\vec{B}_{\rm p}/\mu_{\rm o}$ is the MHD Poynting flux, and ${\rm d}\vec{A}={\rm d}A\vec{n}$. Where ${\rm d}A=2\pi r{\rm d}r$ is the differential area
of the disc surface. Thus, the fraction of the accretion power carried away by the ejecting materials is \citep{Fer97,Foe08b,Petr10}
\begin{equation}
\frac{P_{\rm ejec}}{P_{\rm acc}} = \frac{\Lambda}{1+\Lambda}\Pi,\label{eq:ejec}
\end{equation}
where the parameter $\Pi$ is introduced to measure the fraction of magnetic energy that feeds the ejecting materials versus that being dissipated by the Joule effect within the JED. $\Pi$ is still poorly constrained \citep[see e.g.][]{Fer97}, and is negatively correlated with the disc aspect ratio $\varepsilon$ (\citealt{Fer10}), whereas $\varepsilon$ is associated with the disc temperature by Eq. (\ref{Temp});  in the current work we set $\Pi$ as a free parameter.

\begin{figure*}
\begin{center}
\includegraphics[width=80mm]{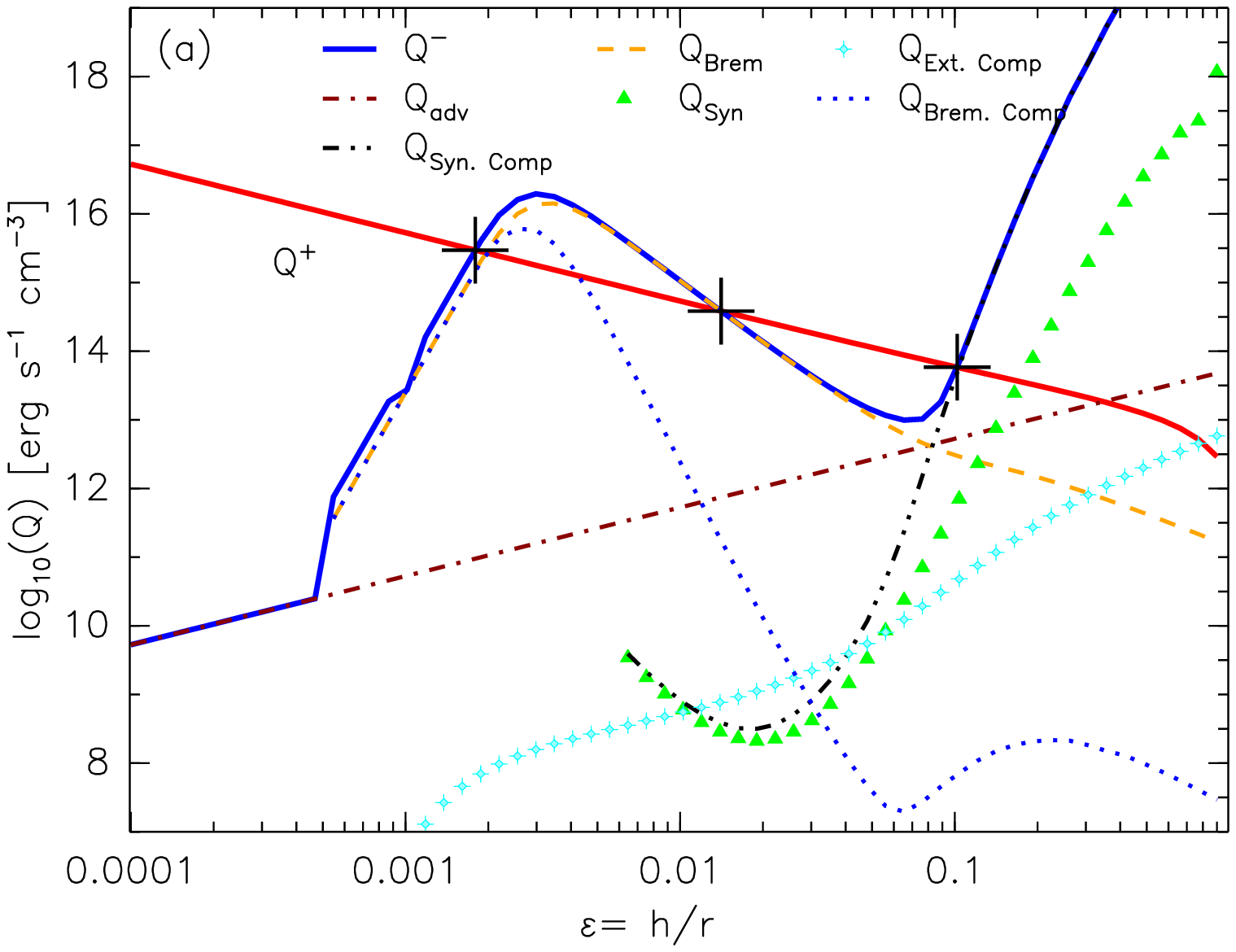}
\includegraphics[width=80mm]{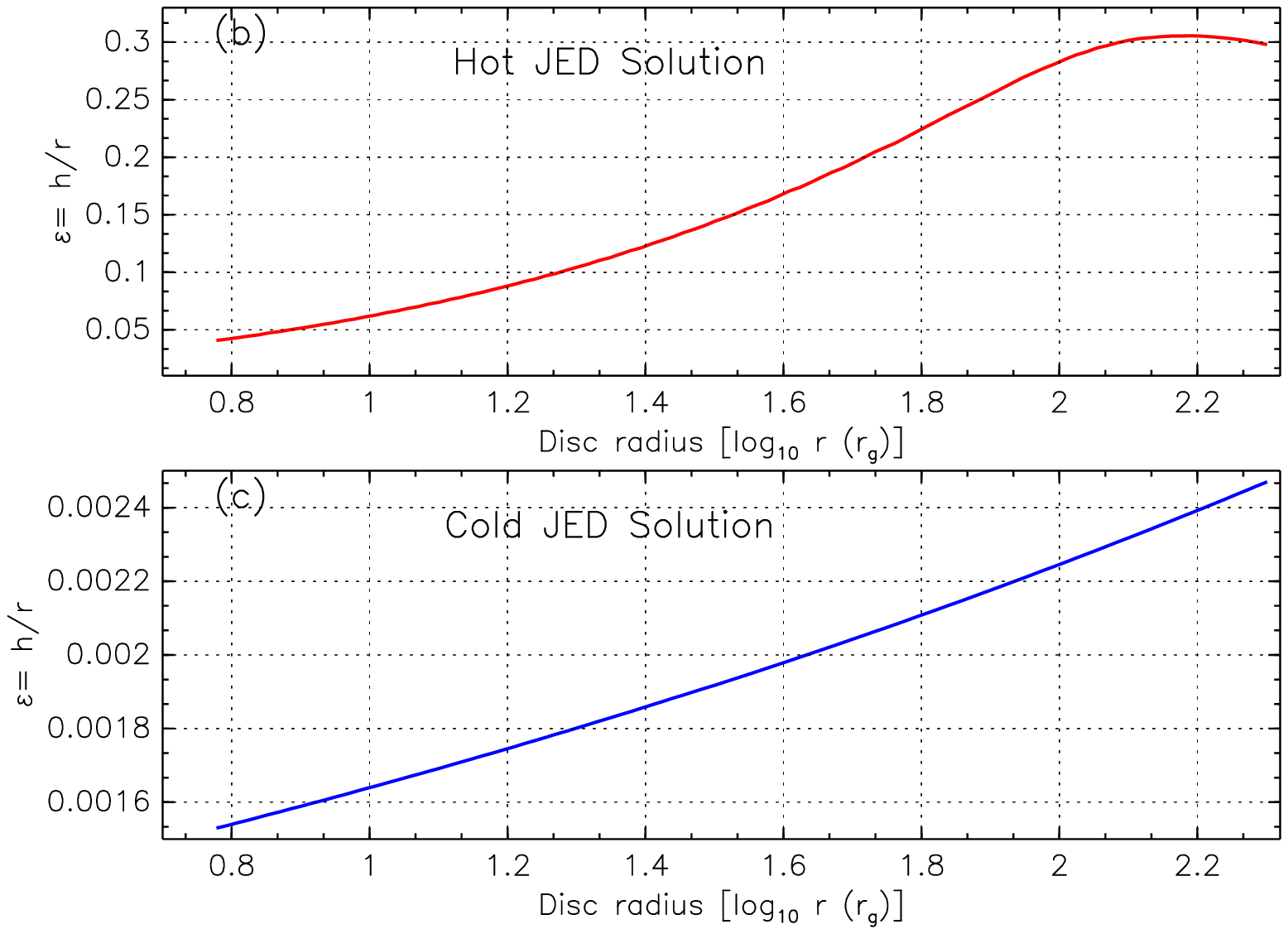}\\
\caption{Left panel (a): the energy balance curve of JED, as a function of the aspect height ratio $\varepsilon$, for a given radius $r=20~r_{\rm g}$. The crosses mark the three thermal equilibrium solutions. Right panel: the aspect height ratio $\varepsilon = h/r$, as a function of the accretion disc radius $r$, for the hot (panel b) and the cold (panel c) solutions. We in this figure adopt the black hole mass $M_{\rm BH}=10~M_{\rm \odot}$, the distance $d=2\ \rm kpc$, the accretion rate at transition radius $r_{\rm tr}=200~r_{\rm g}$ as $\dot{M}_{\rm a,out}= 0.15~\dot{M}_{\rm Edd}$. The ejection efficiency is set as $\xi=0.1$ and $\Pi=0.6$.}
\label{fig:eq}
\end{center}
\end{figure*}

\subsection{JED energy balance}
The detailed structure of the magnetized accretion disc in the JED model depends crucially on the balance between the heating and cooling rates (per unit volume),
\begin{equation}
Q^{+}(r) = Q^{-}(r) = Q_{\rm rad}(r)+Q_{\rm adv}(r).
\label{EQEQ}
\end{equation}
The JED heating term $Q^{+}$ can be derived from the radial differentiation of $P_{\rm JED}$ (note that $P_{\rm JED} = P_{\rm acc}-P_{\rm ejec}$, see Eq. [\ref{energy}]),
\begin{equation}
Q^{+}(r)= \left(1-\frac{\Lambda}{1+\Lambda} \Pi \right) \times \frac{G M \dot{M}_{\rm a}}{8 \pi h r^3}.
\label{Qplus}
\end{equation}
It has been shown that $\Lambda$ is roughly equal to the inverse disc aspect ratio (see \citealt{Fer06,Com08}), $\Lambda \approx 1/\varepsilon$ (ref. Eq. [\ref{LMAD}]). We thus adopt $\Lambda = 1/\varepsilon$ in this paper.

According to the second law of thermodynamics, the advection of entropy can be explicitly expressed as
\begin{equation}
Q_{\rm adv}(r) = Q_{\rm int}(r) -Q_{\rm com}(r),
\label{Qadv}
\end{equation}
where $Q_{\rm int}=\frac{n\upsilon_{\rm r}k_{\rm B}}{\gamma-1}\frac{\partial T}{\partial {\rm r}}$ is the internal energy gradient term (gas adiabatic index $\gamma=5/3$), and $Q_{\rm com}= \upsilon_{\rm r}c_{\rm s}^2\frac{\partial \rho}{\partial r}$ is the compression work.

Provided the radiative cooling is also given (see Sec. \ref{radcool} for details on radiative mechanisms included in the JED model), we can numerically solve Eq.(\ref{EQEQ}) to derive the energy equilibrium of the JED. We note that the basic properties of JED, both dynamical and radiative, are closely related to the temperature of the magnetized accretion disc, thus $\varepsilon$ (ref. Eq. [\ref{Temp}]).

\subsection{JED radiative cooling}
\label{radcool}
In the optically thick case, we use the classical formula of the multi-temperature blackbody radiation (e.g., \citealt{Frank02,Kato08}). In the optically thin case, JED can exhibit various radiative cooling processes, due to the complex environment it embeds. We consider synchrotron and its self Comptonization, bremsstrahlung and its Comptonization, and external Compton scattering of the thermal radiation from the outer SSD. The total radiative cooling in Eq. (\ref{EQEQ}) can be expressed as
\begin{equation}
Q_{\rm rad} = Q_{\rm Brem}+ Q_{\rm Syn} + Q_{\rm Syn. Comp} +Q_{\rm Brem. Comp} + Q_{\rm Ext. Comp}.
\end{equation}

Here we follow \cite{Esin96} for the bremsstrahlung radiation $Q_{\rm Brem}$, \cite{Mah96} for the synchrotron emission $Q_{\rm Syn}$. The Comptonization for various sources of seed photons are also included. For the external Comptonization $Q_{\rm Ext. Comp}$ where seed photons come from SSD, we assume for simplicity that half of the photons from the multi-temperature blackbody radiation of the outer ($\ga r_{\rm tr}$) SSD \citep{Frank02,Kato08} will propagate into the inner ($\la r_{\rm tr}$) JED region, and scatter (Compton up-scatter) with the hot electrons there.

In the current work, we simplify the Compton scattering calculation by adopting the enhancement factor $\eta$ introduced in \cite{Dermer91}. We also assume the spectrum of a power-law type with a Wien bump at its high energy end. Detailed description on the calculation of Compton scattering can be found in \citet{Wardz00}.

Although our methodology for the Compton scattering part is simple thus crude, we find that, at least for the accretion regime explored in this paper, our treatment is roughly consistent with more elaborate models, see Sec. \ref{sec:appl} for comparisons.

\section{Numerical Results}

\subsection{Thermal equilibrium curve}
\label{sec:theq}

In this section, we analyze the JED thermal equilibrium curve. For this purpose, we take the black hole mass to be $M_{\rm BH}=10~M_{\rm \odot}$, a distance to be $2\rm\ kpc$, the transition radius to be $r_{\rm tr}=200~r_{\rm g}$, and the accretion rate at $r_{\rm tr}$ to be $\dot{M}_{\rm a,out}= 0.15~\dot{M}_{\rm Edd}$. We further fix the ejection efficiency $\xi=0.1$ and $\Pi=0.6$. The results are shown in Fig.\ \ref{fig:eq} (a) for a given radius $r=20~r_{\rm g}$. As marked in crosses, three solutions are found. In these solutions, both the cold and hot solutions are thermally stable, while the intermediate solution is thermally unstable (see \citealt{SLE76} for related discussions).

As shown in Fig.\ \ref{fig:eq}(a), the cooling components that contribute to the hot equilibrium solution are advection, synchrotron and its Comptonization, and external Comptonization. On the other hand, only bremsstrahlung and its Comptonization (actually blackbody radiation due to high optical depth, see below) are responsible for the cold equilibrium solution. We further plot in Fig.\ \ref{fig:eq}(b\&c) the aspect height ratio $\varepsilon$ as a function of radius $r$. The aspect ratio is positively correlated to the radius for both the hot and the cold solutions. However, the saturation (or slight turnover) of $\varepsilon=h/r$ for the hot solution near $r_{\rm tr}$ is due to the fact that the dominating cooling is the advection. As for the same parameter values (e.g. $\xi$, $\Pi$ and $\dot{M}_{\rm a,out}$), if neglecting contributions of radiative cooling, we are able to get the maximum value of $\varepsilon\simeq 0.31$ at near $r_{\rm tr}$ by equating the expression of $Q^{+}(r)=Q_{\rm adv}(r)$. We note that, this positive correlation is also shown in SSD, with $\varepsilon \propto r^{1/8}$ \citep{Frank02}. Through numerical fitting, we find that our cold branch of JED has $\varepsilon \propto r^{0.137}$ for current adopted parameters, with the index slightly larger than that of SSD. We note that the positive relationship between $\varepsilon$ and $r$ is crucial for the irradiation model of SSD, where the radiation from inner regions can heat the outer regions. A similar situation is expected to happen for the cold JED solution as well, and the irradiation effect should be slightly stronger.

We now examine the energy power for the solutions shown in Fig.\ \ref{fig:eq}. As defined in Eq.\ (\ref{EQEQ}), the cooling includes radiation and advection, and the heating is the total liberated accretion power. From Eq. (\ref{Qplus}), we find that, for given $\dot{M}_{\rm a}$ and $\Pi$, the hot JED solution (with larger $\varepsilon$) will have larger heating power ($\propto 4\pi r^2 h Q^{+}$), compared to the cold one. The radiative power for the cold branch is $\approx 4.73\times 10^{36}~\rm erg~s^{-1}$, the same to the heating power. Advection is totally negligible here. The hot branch, on the other hand, is rather complicated. The total heating power is $\approx 6.23\times 10^{36}~\rm erg~s^{-1}$. About 86.9 $\%$ of this power, $\approx 4.55\times 10^{36}~\rm erg~s^{-1}$, is liberated as radiation, while the rest 13.1~$\%$ (with the power $\approx 6.79\times 10^{35}~\rm erg~s^{-1}$), is advected inward. This is also different from the ADAF solution, where a much higher fraction, $\lesssim 1$, of the heating power is advected inward \footnote{Note that although ADAF model includes outflow, it does not take $P_{\rm ejec}$ into account in its energy balance. But see \citet{xy08}.}. Although part of the energy is advected inward, we find that the hot branch has comparable (but slightly smaller) luminosity to that of the cold one. To understand this, with Eq. (\ref{Qadv}) above for the advection term, and the dynamical structure given through Eqs. (\ref{Uvel})- (\ref{numDen}), and take $\varepsilon$ to be an ``averaged'' representative value, we have $P_{\rm adv}\approx \varepsilon^2 P_{\rm acc}$ (see also \citealt{Petr10}). For the innermost regions where most of the radiation comes from, e.g. $r\lesssim 40\ r_{\rm g}$, the hot branch has $\varepsilon\lesssim0.15$, implying that advection is insignificant to the energy budget $P_{\rm JED}$ (Note $P_{\rm JED}=(1-\Pi/(1+\varepsilon))P_{\rm acc}$, we thus have $P_{\rm adv}/P_{\rm JED} \approx \varepsilon^2/(1 - \Pi/(1+\varepsilon))$.). Besides, the unimportance of advection in JED also implies that the temperature of JED decreases with decreasing radius $r$ in inner disc region due to strong radiative losses (see Fig.\ \ref{fig:4panels}b).

\begin{figure}
\begin{center}
\includegraphics[width=80mm]{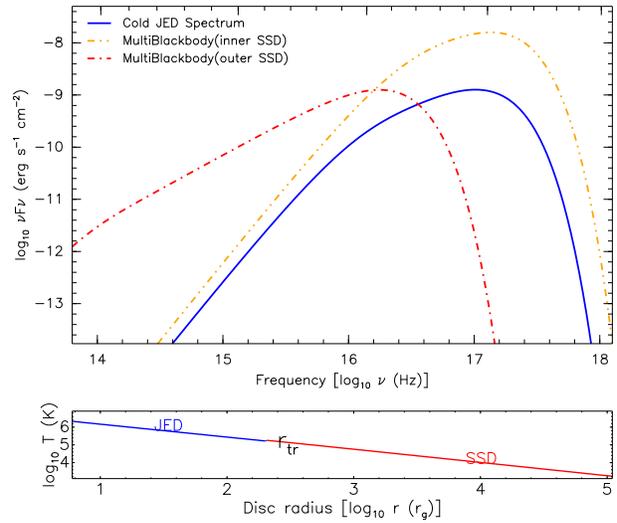}
\caption{Top panel: the spectrum of the cold branch of JED models. The bremsstrahlung (together with its Comptonization) dominates the radiative cooling of inner JED. For comparison, we also show in dot-dashed curve the blackbody radiation from the outer SSD, and in the dot-dot-dashed curve the blackbody radiation if the inner disc is an SSD. Bottom panel: the temperature distribution (as a function of radius) of the whole accretion disk (JED + SSD). Model parameters are the black hole mass $M_{\rm BH}=10~M_{\rm \odot}$, the distance $d=2\ \rm kpc$, the accretion rate at transition radius $r_{\rm tr}=200~r_{\rm g}$ as $\dot{M}_{\rm a,out}= 0.15~\dot{M}_{\rm Edd}$. The ejection efficiency is set as $\xi=0.1$ and $\Pi=0.6$..}
\label{fig:cold}
\end{center}
\end{figure}

\begin{figure}
\begin{center}
\includegraphics[width=80mm]{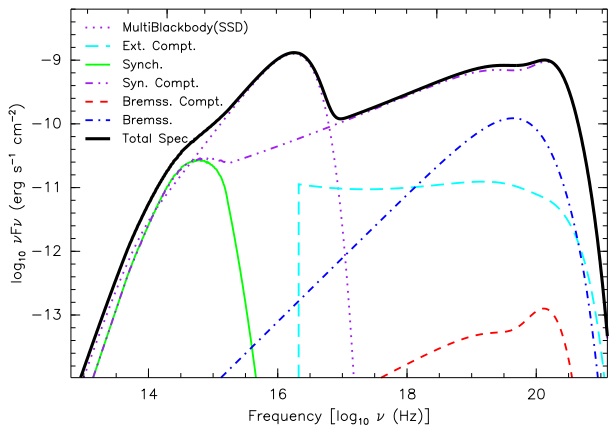}
\caption{ The SEDs of the hot branch of JED models. Various contributions are included, i.e. synchrotron, bremsstrahlung, and their Comptonization (see labels in the figure). The adopted parameters are the black hole mass $M_{\rm BH}=10~M_{\rm \odot}$ at a distance $2\ \rm kpc$, the accretion rate at transition radius $r_{\rm tr}=200~r_{\rm g}$ as $\dot{M}_{\rm a,out}= 0.15~\dot{M}_{\rm Edd}$. The ejection efficiency is $\xi=0.1$ and $\Pi=0.6$.}
\label{fig:spec}
\end{center}
\end{figure}

\subsection{Spectral energy distributions}
\subsubsection{Cold solution}

Now we investigate the spectral properties of JED models. Basic parameters are those in Sec. \ref{sec:theq}. We first in Fig.\ \ref{fig:cold} present the theoretical spectrum of the cold solution branch of the JED model. Since the temperature is very low, synchrotron is negligible due to the lack of relativistic electrons and the weak magnetic field. The main radiation comes from bremsstrahlung and its Comptonization; in fact, under large optical depth, the spectrum will be multi-blackbody.

However, we should emphasis that the cold branch of the JED model is different from SSD in several aspects. First, the radial velocity of JED is comparable to the sound speed, while the SSD is highly subsonic. Correspondingly, the optical depth of JED is much smaller than that of SSD with the same accretion rate. In our calculation, the optical depth of cold JED solution is $\sim 4$ and $\sim 25$ at radii $r_{\rm tr}$ and $r_{\rm in}$, respectively. Second, the temperature profile of cold JED is different from that of SSD \citep{Cas00b}, especially for cases with larger ejection efficiency $\xi$. Third, a large fraction of the accretion power is used to power the ejections, i.e. $P_{\rm ejec}\approx \Pi P_{\rm acc}\approx 0.6 P_{\rm acc}$ (ref. Eq.\ [\ref{eq:ejec}]) in the JED model, for our adopted parameters. This, together with lower liberated accretion power for the JED model because of mass loss (see Eq.\ [\ref{eq:Pacc}]), implies that our JED models have much lower luminosity compared to SSD with the same accretion rate at outer boundary (see Fig.\ \ref{fig:cold}). Neglecting advection, which is reasonable at such low temperature and disc aspect ratio, we have
\begin{eqnarray}
P_{\rm rad,JED}\approx P_{\rm JED} &=& P_{\rm acc,JED}-P_{\rm ejec,JED} = (1-\frac{\Pi}{1+\varepsilon}) P_{\rm acc,JED} \nonumber \\
 &\approx& 0.4 P_{\rm acc,JED}\approx 0.28 P_{\rm rad,SSD}.
\end{eqnarray}
In the above expression, $P_{\rm rad,SSD}$ is the expected radiative emission of an SSD assuming the SSD extends down to $r_{\rm in}$ and radiates all its liberated accretion power, i.e. $P_{\rm rad,SSD}=P_{\rm acc,SSD}$ \citep{Frank02, Kato08}. We also assume a constant accretion rate (non-function of the radius) for the SSD model (mass loss efficiency $\xi=0$). For the JED, we assume an ejection efficiency mass loss efficiency $\xi=0.1$.

\subsubsection{Hot solution}
\label{Num:HotS}
In this section, we focus on the hot branch, which is more attractive, and is the main aim of this research. To avoid confusion, JED hereafter only refers to the hot solution of the JED model. We show the spectrum of the hot branch in Fig.\ \ref{fig:spec}. As shown in this figure, the radiation in optical and hard X-ray bands are produced by the JED itself. For our adopted accretion rate, the optical band is dominated by synchrotron radiation, while the (hard) X-ray band ($\sim 10^{17-21}\ {\rm Hz}$) is the self Comptonization of synchrotron radiation. The UV and soft X-ray bands ($\sim 10^{16-17}\ {\rm Hz}$) is dominated by the blackbody radiation from the outer SSD. As for the currently adopted accretion rate (relatively high), bremsstrahlung and its Comptonization are negligible in JED models. One important conclusion is that JED shares similar radiative mechanisms to the ADAF model at lower accretion rate or the LHAF model at higher accretion rate (e.g., \citealt{Yuan05}), with the difference that it self-consistently includes ejection. The JED can thus produce a spectrum applicable to the hard and quiescent states of X-ray binaries, and low-luminosity active galactic nuclei, which is the field where hot accretion flow models (ADAF or LHAF) are successfully applied \citep{Yuan07,NM08}.

\begin{figure*}
\begin{center}
\includegraphics[width=120mm]{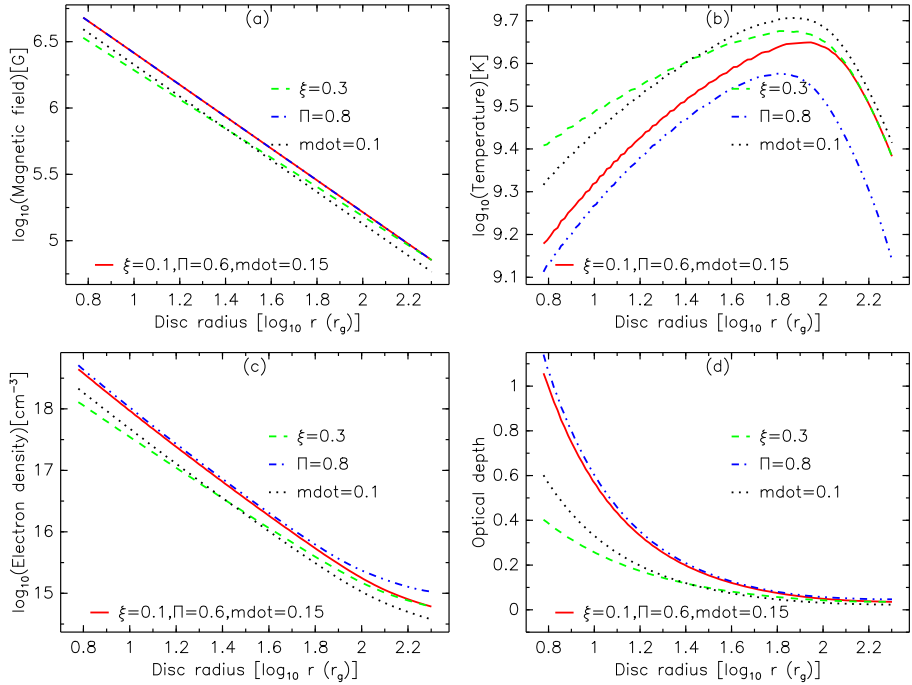}
\caption{The dynamical structure, i.e. of the magnetic field (panel a), electron temperature (panel b), electron number density (panel c) and optical depth (panel d) for JED models. The solid lines in each panel are considered as
our benchmark (The used parameters are the same as that of Fig. \ref{fig:spec}). As marked in each panel, we also illustrate the dynamical structure changes due to the modification in outer boundary accretion rate $\dot{M}_{\rm a,out}$ (dotted lines), in ejection efficiency $\xi$ (dashed lines) and $\Pi$ (dot-dot-dashed lines).}
\label{fig:4panels}
\end{center}
\end{figure*}

To help to understand the SED properties of JED models, we plot in Fig.\ \ref{fig:4panels} the corresponding dynamical structures, i.e. the magnetic field, the electron temperature and number density, and the optical depth (see solid lines in each panel), for the JED models shown in Fig.\ \ref{fig:spec}. After setting those parameters (corresponding to solid lines) as the fiducial values, we then change outer boundary accretion rate into $\dot{M}_{\rm a,out}=0.1$ (dotted lines), or ejection efficiency into $\xi=0.3$ (dashed lines) and $\Pi=0.8$ (dot-dot-dashed lines). As shown in this figure, with increasing the ejection efficiency $\xi$, we find that the JED model (dashed lines) inclines to have higher electron temperature (Fig.\ \ref{fig:4panels}[b]), lower magnetic field (Fig.\ \ref{fig:4panels}[a]), electron number density (Fig.\ \ref{fig:4panels}[c]) and optical depth (Fig.\ \ref{fig:4panels}[d]). Similarly, we can obtain the almost same results if a lower accretion rate is set. Here, small optical depth can be explained by $\tau\equiv n \sigma_{\rm T}h\propto \dot{M}_{\rm a}/v_{\rm r}$, where $\sigma_{\rm T}$ is the Thompson cross section. Higher electron temperature and lower optical depth imply a Compton inefficient process of radiation. As shown in Fig.\ \ref{fig:4panels}[a], changes in the value of $\Pi$ does not change the magnetic field, thus changes of synchrotron radiation spectral shape are mainly affected by electron properties rather than the magnetic field for the same accretion rate (see also Eq. [\ref{MagField}] and Sec.\ref{Num:Par}). The increase of $\Pi$ will result in slightly increasing electron number density (Fig.\ \ref{fig:4panels}[c]) and optical depth (Fig.\ \ref{fig:4panels}[d]), but significantly decreasing electron temperature (Fig.\ \ref{fig:4panels}[b]). In this model, most of the radiation comes from the inner regions where JED has a large Compton y-parameter ($y \equiv \frac{4kT_{\rm e}}{m_{\rm e} c^2}\tau \propto T_{\rm e}\tau$), we thus expect JED to produce a strong Comptonized radiative process, and have a hard X-ray spectrum (see Fig.\ \ref{fig:spec}).

We emphasis that the above argument is limited to the single-temperature assumption of JED. It is interesting to re-investigate this problem when the two-temperature JED model is available.

\subsubsection{SED dependency on model parameters}
\label{Num:Par}
\begin{figure*}
\begin{center}
\includegraphics[width=160mm]{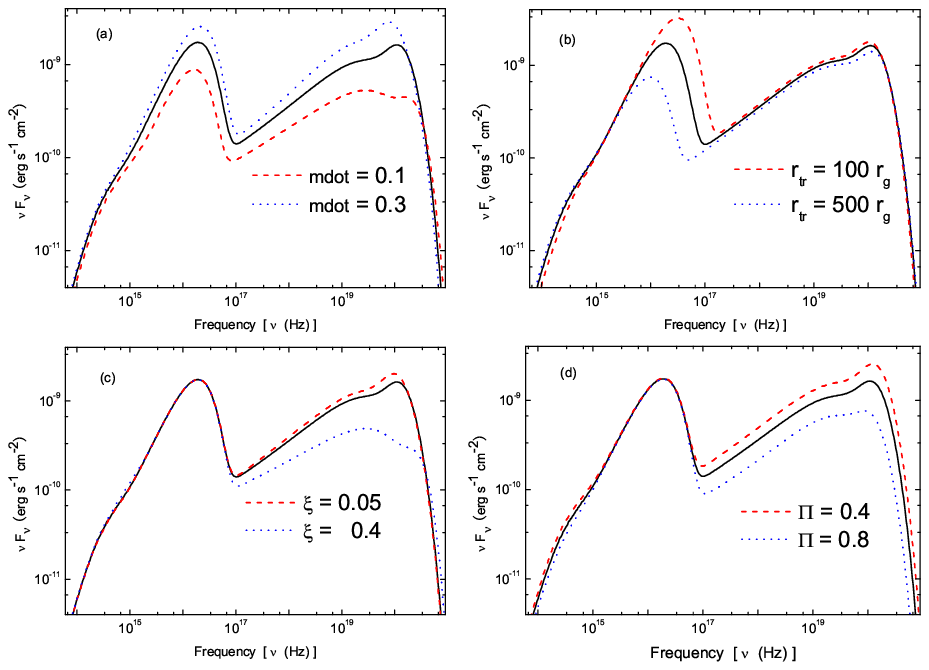}
\caption{Hot JED spectra under various model parameters. In all the four panels, the solid curves represent the emission spectra for base parameters, i.e. the transition radius $r_{\rm tr}=200~r_{\rm g}$, the boundary accretion rate $\dot{M}_{\rm a,out}= 0.2~\dot{M}_{\rm Edd}$, and the ejection efficiency $\xi=0.1$ and $\Pi=0.6$. As marked in each panel, we illustrate the spectral change due to the modification in outer boundary accretion rate $\dot{M}_{\rm a,out}$ (panel a), in transition radius $r_{\rm tr}$ (panel b), in ejection efficiency $\xi$ (panel c) and $\Pi$ (panel d).}
\label{fig:JEDmodels}
\end{center}
\end{figure*}
Before reproducing the spectrum for individual sources, we first study the spectral modifications due to the change of various model parameters. Those parameters include the outer boundary accretion rate (at transition radius) $\dot{M}_{\rm a,out}$, the transition radius  $r_{\rm tr}$, the ejection efficiencies $\xi$ and $\Pi$. As shown in Fig. \ref{fig:JEDmodels}a, the increase of accretion rate will lead to increase of fluxes in the UV band ($\sim 10^{16}-10^{17}$~Hz; radiation from outer SSD) and X-ray band (Comptonized radiation from inner JED). Also, the X-ray spectrum will be harder, mainly because of the increase in Compton y-paramter.

We then fix the boundary accretion rate, and vary the transition radius. The results are illustrated in Fig. \ref{fig:JEDmodels}b. Larger $r_{\rm tr}$ leads to several consequences. First, the peak frequency of the multi-blackbody radiation, which is determined by the temperature at $r_{\rm tr}$, is reduced. Second, the total flux in the UV band is also reduced. Third, the effective accretion rate for the JED will also be reduced (ref. Eq. \ref{EQmot}), leading to lower flux in X-ray and optical bands. Finally, the X-ray spectrum is also slightly softer, as explained above.

For fixed boundary accretion rate and transition radius, the outer SSD is determined, thus the spectrum of this component (in $\sim 10^{16}-10^{17}$~Hz) is fixed. We now investigate the consequences of varying $\xi$ and $\Pi$. $\xi$ characterizes the JED mass ejection efficiency. Larger $\xi$ is similar to the reduction of accretion rate in JED. The corresponding spectral changes (shown in Fig. \ref{fig:JEDmodels}c) can be easily understood. We should also point out that, changes in either $\xi$ or $\dot{M}_{\rm a,out}$ will lead to a modification in the accretion rate within JED regime. However, there are several differences between them. Varying $\xi$ modifies the relative contribution of radiation from different radii, thus modifies the spectral shape. Besides, the change in $\dot{M}_{\rm a,out}$ has its consequence in the radiation from outer SSD, while the change in $\xi$ does not.
Finally, as clearly illustrated in Fig. \ref{fig:JEDmodels}d, larger $\Pi$ means that more energy will be deposited in the ejecting material (see also Sec. \ref{Num:HotS}, for the dynamical properties). Consequently, the infrared-optical ($\sim 10^{14}-10^{16}$~Hz) and hard X-ray fluxes will be lower. The spectrum of X-ray is slightly softer.

\section{APPLICATION TO THE HARD STATE OF X-RAY BINARIES: the case of XTE J1118+480}
\label{sec:appl}

From above, the hot solution of the JED model can reproduce the spectrum for the hard state of black hole X-ray binaries. Below we focus on the microquasar XTE J1118+480.

\begin{figure}
\begin{center}
\includegraphics[width=80mm]{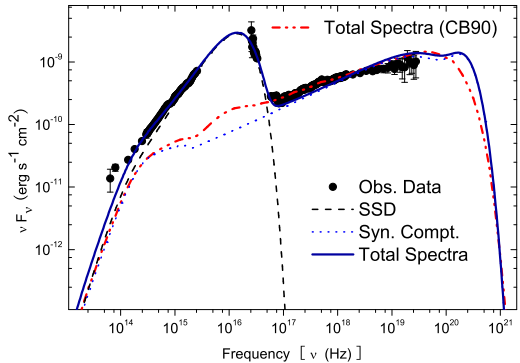}
\caption{ The SED of the microquasar source XTE J1118+480 in JED model. The fitting parameters are $r_{\rm tr}=500~r_{\rm g}$, $\dot{M}_{\rm a,out}= 0.72~\dot{M}_{\rm Edd}$, $\xi=0.5$ and $\Pi=0.48$. Observational data correspond to the March 2000 outburst, which is taken from \citet{Mai09}.}
\label{fig:1118}
\end{center}
\end{figure}

XTE J1118+480 was first observed in March 2000 by the {\it RXTE} telescope. During the past years, this source exhibits several outbursts, during which conical hard state is obvious. One compelling advantage of this source is that, it lies well above the Galactic plane ($b\approx63$ deg), thus remains observable in optical and near-infrared bands due to very low reddening. During the hard state, the observations in these bands can reveal the blackbody radiation from the outer SSD, thus help to constrain the transition radius $r_{\rm tr}$. For our investigations here, we take its black hole mass to be $M_{\rm BH} = 8.5~M_{\rm \odot}$ and the distance to be $d=1.72~{\rm kpc}$ \citep{Gelino06}. The broad-band quasi-simultaneous observational data is collected by \cite{Mai09}, and we use its 2000 outburst (circular points in Fig.\ \ref{fig:1118}), which is also investigated in jet-accretion model by \citet{Yuan05} and in jet model by \citet{Mark01} and \citet{Peer12}. The radio observation, which originates from the relativistic jet, is not included here for simplicity (see e.g., Fig. 2 in \citealt{Yuan05}).

We show in Fig.\ \ref{fig:1118} a fit obtained with the JED model. In this fit, the transition radius is found to be $r_{\rm tr}=500~r_{\rm g}$ and the accretion rate at $r_{\rm tr}$ to be $\dot{M}_{\rm a,out}= 0.72~\dot{M}_{\rm Edd}$. Furthermore we find the ejection efficiency $\xi=0.5$ and the parameter $\Pi=0.48$. Our spectral fitting is basically satisfactory. Although our calculation in the Compton scattering is simplified, a comparison to the result reproduced by a widely-tested and elaborate Compton scattering code (dot-dot dashed curve in Fig.\ \ref{fig:1118}, which is based on the work by \citealt{Coppi90}, see \citealt{Yuan05} for more details) indicates that our treatment is roughly valid, at least for our application to XTE 1118+480 here.

The spectral fitting of JED to the observational data requires a large value of $\xi$. Observationally, this is because, a relatively high outer boundary accretion is required to reproduce the UV bump, while a relatively low JED accretion rate is needed for the X-ray observation. First through the UV bump fitting (from outer SSD), the outer boundary accretion $\dot{M}_{\rm a,out}$ and the transition radius $r_{\rm tr}$ can be fixed in advance, and then we can adjust the parameters $\xi$ and $\Pi$ to fit observations in X-ray bands. Because of adjustment of the parameter $\Pi$, which does not change spectral shape (into a softer spectrum; see aslo Fig. \ref{fig:JEDmodels}[d]), we are only able to adjust the ejection efficiency $\xi$. However, this value lies outside of the suggested parameter range of $\xi$ (e.g. $\lesssim0.15$ in \citealt{Fer97,Cas00a}). We note that when heat deposition occurs at the JED upper layers, the value $\xi \sim 0.5$ can be reached \citep{Cas00b}. In a word, how large can $\xi$ be is an open question, especially if the JED portion of the accretion disc gets irradiated (Jonathan Ferreira, private communication).

We note that jet-accretion model argues the radiation in the optical wave bands to be from synchrotron (see e.g., Fig. 2 in \citealt{Yuan05}) thus highly polarized, while the JED model argues the corresponding radiation to be from SSD (see Fig. \ref{fig:1118}), thus only weak polarization can be observed. At the current stage, the JED model is still a single-temperature structure\footnote{At extremely high accretion rate, the interaction of electron-ions  become very strong mainly due to the Couloumb interactions and the hypothesis of a single temperature is reasonable. As a result, electrons and ions have the same temperature (e.g., \citealt{BKL97,Quat98}).}. However, experience from the development of ADAF indicates that, the coupling between electrons and ions is insufficient (\citealt{Nar95}, see also \citealt{SLE76}), thus the flow is likely to be a two-temperature structure, with ions much hotter than electrons. Investigating the JED structure and correspondingly the spectrum under two-temperature configuration is beyond the scope of this work, and will leave for future study.

\section{DISCUSSIONS AND SUMMARY}
JED models have been developed and applied to study black hole X-ray binaries in their hard state \citep{Fer06,Petr10} and protostellar objects (\citealt{Com08}). In this work, by including various radiative cooling processes (external Comptonization from outer SSD, synchrotron and its Comptonization, and bremsstrahlung and its Comptonization), we reinvestigate the energy balance properties and SEDs of the JED model (for earlier work, see \citealt{Foe08a,Foe08b}). Subsequently we numerically calculated the spectrum for the stable solutions (both the cold and the hot branches). We also make a comparison of the cold solution to the widely adopted SSD model.

After reproducing theoretical SED, we apply this model to the microquasar XTE J1118+480 as an example. We find the hard spectrum of XTE J1118+480 can also be explained by the JED model.

Although the physical interactions between the hot accretion flow and ejections (jet, outflow) are still not fully understood, they are generally considered as some fundamental components of accretion processes. Many works focus on understanding these physical processes at different accretion rates. One previous research speculates that there are some common mechanisms responsible for the accretion-ejection connections for all accreting black holes (\citealt{Zhang07}). Recently, MHD numerical simulations from a hot accretion flow model support the existence of powerful accretion disc winds (\citealt{TN11,MT12}). From observations, there are also some investigations supporting the presence of winds (e.g., \citealt{Lee02, D11,Ponti12}); these winds have an equatorial geometry with opening angles of few tens of degrees (\citealt{Ponti12}). In the framework of JEDs, some further steps on the dynamical properties and radiative features of disc wind are necessary (see also \citealt{Fer13}).

One evident advantage of the JED model is that, it conserved both mass and energy. Part of the accreting material is simultaneously redistributed as ejections self-consistently. Correspondingly, these ejections will also carry away part of the accreting energy (ref. Eq. [\ref{energy}]). The disadvantage is that, compared to ADAF model, currently the JED model is still a single-temperature one. However, like ADAF, the coupling of electrons and ions in JED is also expected to be weak, thus a two-temperature structure will be formed naturally. In this sense, due to lower electron temperature, the luminosity of the hot JED will be lower than the single-temperature JED model here. We leave this for future investigations.

\section*{ACKNOWLEDGMENTS}
We appreciate the anonymous referee for his/her careful reading of our manuscript, and valuable comments and suggestions that significantly improve the demonstration of our work. It is a pleasure to thank Dr. C\'{e}dric Feollmi for providing his energy balance code. We appreciate Dr. Feng Yuan for his encouragement and valuable comments on an earlier draft,  Drs. Asaf Pe'er and Dipankar Maitra for providing us the observational data. JFZ is partly supported by the Guizhou provincial Natural Science Foundation (No. 2010080) and the Science and Technology Foundation of Guizhou Province (Nos. J[2012]2314 and LKT[2012]27). FGX is supported by the Natural Science Foundation of China (No. 11203057) and the SHAO key project (No. ZDB201204)


\end{document}